\documentclass[12pt,a4paper]{article}
\usepackage[hyperref]{emnlp2020}
\usepackage{times}
\usepackage{latexsym}

\usepackage{graphicx, float, multirow, enumitem}
\graphicspath{ {images/} }

\usepackage{microtype}

\aclfinalcopy 

\title{LonelyText: A Short Messaging Based Classification of Loneliness}

\author{Mawulolo K. Ameko, Sonia Baee, Laura E. Barnes \\
  University of Virginia \\
  \texttt{mka9db,sb5ce,lb3dp@virginia.edu} 
}

\begin{document}
\maketitle

\begin{abstract}
\normalsize
Loneliness does not only have emotional implications on a person but also on his/her well-being. The study of loneliness has been challenging and largely inconclusive in findings because of the several factors that might correlate to the phenomenon. We present one approach to predicting this event by discovering patterns of language associated with loneliness. Our results show insights and promising directions for mining text from instant messaging to predict loneliness.
\end{abstract}


\section{Introduction}
Loneliness isn't just a fleeting feeling, leaving us sad for a few hours to a few days. Research in recent years suggests that for many people, loneliness is more like a chronic ache, affecting their daily lives and sense of well-being. Whether briefly acknowledging a passerby, conversing with a friend over a cup of coffee, or sending quick text messages to friends and colleagues, frequent social interaction in many ways defines the human experience. 

\subsection{Problem Definition}
With the advent of social media, we have more access to others than ever before, so why does loneliness continue to be a pervasive social problem?
Despite lacking the palatability of in-person social interactions, psychological research suggests that social media interactions should elicit very real psychological consequences. Minor similarities such as knowing that someone happens to like the same painting during a laboratory study can lead to feelings of affiliation and a preference for subsequently socializing with someone~\cite{tajfel1974familiarity}. Conversely, even the experience of being excluded from a computer game of catch between two cartoon characters can elicit the hurt feelings and affiliative desires of real social rejection \cite{williams2000cyberostracism}. On a larger scale, loneliness and rejection are public health issues. Living alone increases the mortality rate of seniors by over 20\%, which is comparable to smoking and drinking, and higher than obesity. Loneliness is also associated with negative health outcomes, including heart disease~\cite{holt2015loneliness, holt2010social}.
Different modes of communication via social media~(e.g., likes vs. posts) can be differentiated from each other, as well as from face-to-face conversation, based on the presence or absence of social cues. If the absence of cues, such as facial expressions or real-time responding, makes social media-based interactions less fulfilling, then social media should have negative effects on well-being and one\textquotesingle s sense of belonging, as some past studies have found~\cite{kross2013facebook}. But, an individual\textquotesingle s experience of the social world depends on how the individual interprets social situations rather than on objective aspects of a situation~\cite{shaked2017breaking}. To the extent that people fill in information, for example, by imagining someone\textquotesingle s tone, based on their expectations, the psychological effects of virtual interactions should vary by person and situation.

In this work, we set out to investigate the question of how a person's language pattern might be characteristic of loneliness. Should this problem be more amenable to topic modeling, then, are lonely people more likely to engage in some specific topics of discussions and then can we find a generalizable pattern? The main contribution of this work is in the leveraging of text mining techniques to provide new insights in the study of loneliness and human well-being.  
The results will ultimately support clinicians to deploy scalable interventions to improve patients' social interaction or help them when they feel lonely. For example, send their favorite contacts some notifications to engage a patient in a conversation along a certain line.

\section{Related works}

Previous work has shown conflicting conclusions with respect to the use of social media. A study conducted among undergraduate students from a university in the US showed that the use of social media leads to more loneliness~\cite{kim2009loneliness} while others show exactly the opposite~\cite{moody2001internet},~\cite{whang2003internet}. The latter study was not conclusive but proposed a further study to corroborate the findings. 
One of the studies has a hypothesis that the relationship between loneliness and preference for online social interaction is spurious \cite{caplan2006relations}.
In another study they hypothesis that user\textquotesingle s personality traits may be crucial factors leading them to engage in participatory media. The literature suggests factors such as extraversion, emotional stability, and openness to experience are related to uses of social applications on the Internet. Results revealed that while extraversion and openness to experiences were positively related to social media use, emotional stability was a negative predictor, controlling for socio-demographics and life satisfaction. These findings differed by gender and age. While extroverted men and women were both likely to be more frequent users of social media tools, only the men with greater degrees of emotional instability were more regular users. The relationship between extraversion and social media use was particularly important among the young adult cohort. Conversely, being open to new experiences emerged as an important personality predictor of social media use for the more mature segment of the sample~\cite{correa2010interacts}, based on these types of studies we can see that the personality of users has a relationship with social media usage.
Users who consume greater levels of content report reduced bridging and bonding social capital and increased loneliness~\cite{burke2010social}. In another study, they found that internet use was found to decrease loneliness and depression significantly, while perceived social support and self-esteem increased significantly~\cite{shaw2002defense}. A study conducted among college students in Hong Kong shows a worrisome vicious cycle between loneliness and Internet addiction~\cite{yao2014loneliness}.
One study was checking the relationship between posting Facebook status update and loneliness. They found that increase in status updating activity reduced loneliness, that the decrease in loneliness was due to participants feeling more connected to their friends on a daily basis, and that the effect of posting on loneliness was independent of direct social feedback (i.e., responses) by friends~\cite{deters2013does}.
Other studies try to capture public mood and emotion by using social media, in one study they found that events in social, political, cultural and economic sphere do have a significant, immediate and highly specific effect on the various dimensions of public mood~\cite{bollen2011modeling}. Some study refers to social media data and internet usage as a digital footprint which can be used to predict personality traits. These social media-based predictions can then be used for a variety of purposes, including tailoring online services to improve user experience, enhance recommender systems~\cite{ameko2020offline}, and as a possible screening and implementation tool for public health. They found that accuracy of predictions is consistent across Big 5 traits, and that accuracy improves when analyses include demographics and multiple types of digital footprints~\cite{azucar2018predicting}. In another study, they have compared different systems for personality recognition from text on a common benchmark. The results show that personal recognition is a challenging task, due to the fact that there are no strong predictive features, nor rather they are very sparse~\cite{celli2013workshop}.
In some new works, they found that basic emotions provide considerable insights in identifying twitter users who suffer from depression. Besides, additional information can be discovered by analyzing these features over time~\cite{chen2018mood}. We will approach this problem differently by studying language patterns that relate to loneliness.

\section{Experimental Methodology}
The study design includes university students for two main reasons: (a) there is a high rate of loneliness among young adults; and (b) recruiting young adults in a university setting will provide a relatively homogeneous sample in terms of life phase, psycho-social stressors, and life experiences, thereby eliminating a wide variety of potential nuisance factors.

As an experience sampling study, the study was conducted over the course of 2 weeks (which participants be alerted at various times during the day on their smartphones to fill out a short questionnaire), and will also involve two lab visits. Participants show up at the psychology lab at the beginning and end of the study. Participants also answer questions through a phone application over the course of their day.

\subsection{Experience Sampling (14-Day Period)}
At the end of Lab Session 1, participants were asked to install data tracking applications on their smartphones. Over a period of 14 days, the software (i) tracked communication statistics, such as time of day, and the number and duration of calls and texts that are made and received; (ii) prompted participants at 3-5 random times each day to answer short items about their current mood, thoughts, behavior, and location/activity (e.g., hanging out at home, attending class, social gathering);
(iii) reminded participants to answer a slightly more comprehensive set of items related to emotions and activities before they go to bed and (iv) tracked the participants' location and motion activity using GPS and accelerometers sensors on the smartphones.

The data used in this study was collected as part of a larger effort to form a comprehensive multimodel dataset that includes personal communication~(i.e., SMS, and call history), social media data~(i.e. Facebook), and Ecological Momentary Assessment~(EMA) survey. The study protocol was approved by the Social and Behavioral Sciences Institutional Review Board~(IRB) at the University of Virginia~(UVa).

The EMA survey was triggered six times in window of two hours as a Random Time (RT) which was asking question related to their current feeling about loneliness. One EMA was triggered at 10 pm of every day during the study as an End Of Day~(EOD) which the user should answer based on his/her feeling during whole that day.
We have 160 participants who they responded to the EMA survey. The average number of response to the RT was 11 times and for EOD 4 times during the study. The ages ranges of our participants from 17 to 24~(Mean = 18.75, Standard Deviation = 1.08).


\section{Methods}

\subsection{Text Data and Preprocessing}
To capture the textual information related to loneliness of the participants, we used both SMS and Facebook messaging over the course of the two weeks study. We considered a window of one day for the annotation of the data with the Ecological Momentary Assessment~(EMA) label. The choice of the window allows us to capture more text to create a rich information base for the label. The dataset consisted of ~2000 days of text across all 112 participants in the study. The EMA label for loneliness was on a scale of $[0, 100]$ which we binarized for the classification task. We considered a threshold of 50 as a natural tipoff point for determining lonely versus not-lonely.

The document refers to the each participant's texts in daily window. The documents for each participants were processed by removing the stopwords, stemming, and pruning words that did not have any meaning such as ``nan" which was referring to the empty row in Facebook messages.

Based on our participants data we just have 118 participant's Facebook data which we just used the Facebook messages. We also have 79 participants' SMS data. 
We have tried to map the Facebook messages to the duration of the study which causes to miss some participants who they did not use their Facebook messenger during the study so we ended up with 112 participants. We used a dictionary to map some common instant messaging jargon to standard word. For example, ``lollll and loooolll'' are mapped to ``lol''. We discarded some head words and tail words based on term frequency and included on term that were at least 2 character to avoid using words like ``t'' or ``s'' in our analysis. We also used tf-idftf for our SVM models.

\subsection{LDA model}
The latent Dirichlet allocation~(LDA) model~\cite{blei2003latent} is a generative probabilistic model that uses a small number of topics to describe a collection of documents. The LDA model captures the multi-topical feature of documents and treats a document as a mixture of words from topics represented as multinomial distributions over the vocabulary. For example, when the words ``course" and ``learning" appear in a conversation or document, this words might be used to describe a topic related to studies when on the other hand, the appearance of words like, "girl", "love" or "kissing" might appear in conversations related to romance. Thus conversations related to studies and romantic interests, can be treated as a mixture of words from these two topics. Based on such assumptions, statistical learning techniques can be applied to identify the topics from a collection of documents/conversations. The distributional representation of a concept captures the key relationship between the concept and words: a concept is conveyed by word choice, and the sense of a word depends on context.
\subsubsection*{Model Specification}
Let $C$ = be a set of documents, where $D$ denotes the number of documents in the corpus; a document $d_i = (w_1, w_2, \cdots, w_{Ni})$ consist of a sequence of words; and let w be a word that takes a value from the vocabulary $\{ v_1, v_2, \cdots, v_{V}\}$. Let $T$ be the number of topics of a LDA model and V be the size of the vocabulary of the corpus. The LDA model simulates the generation of a document with the following stochastic process: 
\begin{itemize}
\item For each document, sample a topic proportion vector $\theta = (\theta_1,\theta_2, \cdots, \theta_{T})'$ from a Dirichlet distribution with parameter $\alpha': \theta: \sim Dir(\theta/\alpha)$. This encodes the number for topics that are believed to be contained in the corpus.
\item For each word in the document, sample a topic $z$ according to multinomial distribution governed by $\theta: z \sim Multi(z/\theta)$. This can be seen as assigning a word to a topic.
\item Conditioning on $z$, sample a word $w$ according to the multinomial distribution with the parameter $\theta_z: w \sim Multi(w/\theta_z)$. This corresponds to picking words to represent a concept.
\item The parameter $\theta_t$, with $t \in \{1, 2, \cdots, T \}$ is a $V$-dimensional vector that defines the multinomial word-usage distribution of a topic. It is distributed as Dirichlet with parameter $\beta: \theta \sim : Dir(\theta/b)$.

\end{itemize}

\begin{figure}[h!]
\centering
\includegraphics[width = 0.39\textwidth]{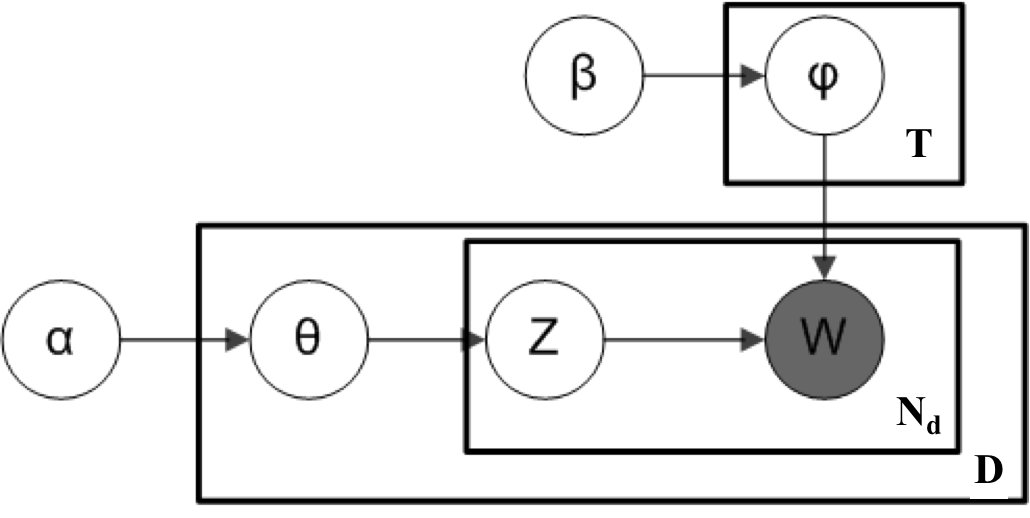}
\caption{The graphical representation of the LDA model. Each node represents a variable, and a shaded node represents an observed variable. Each rectangle plate indicates a replica of the graphical structure. The variables $D$ and $N_d$ at the bottom right of plate indicate the number of replicates of the structure.}
\label{fig:lda_graph}
\end{figure}

The graphical representation of the LDA model is shown in Figure~\ref{fig:lda_graph} in plate notation. The nodes represent random variables, and edges represent the probabilistic relationship, i.e., the conditional probability, between the variables. The shaded and un-shaded nodes represent the observed and unobserved variables, respectively. Each rectangular plate represents a replica of a data structure, e.g., a document; the variable at the bottom right of a plate indicates the number of the replicates. In this graph, each document is associated with a topic composition variable $\theta$, and a total of $N_d$ replicates of topic variable $z$ and word $w$. The graph also shows that there are $T$ topic word distributions.

\subsubsection*{Statistical Inference}
The LDA model extracts the latent semantic topics by estimating the word distributions and assigns each word in a corpus to a latent semantic topic through statistical inference. The exact inference of the parameters and latent variables of a LDA model is intractable. There exist several classical models for approximating the posterior distribution over the parameters such as Markov chain Monte Carlo~\cite{gelfand1990sampling} and Gibbs Sampling~\cite{geman1987stochastic}. In this work, we used variational inference, a method widely used to approximate posterior densities for Bayesian models. The idea behind variational inference is to first posit a family of densities~(e.g. the exponential family) and then to find the member of that family which is close to the target. Closeness is measured by Kullback-Leibler divergence. While the classical methods more precisely approximate the posterior distribution, they are more computationally expensive and thus do not scale as well to growing datasets as variational methods. The reader can refer to~\cite{blei2017variational} for a comprehensive review of variational methods.

\subsubsection*{Support Vector Machine (SVM)}
SVM is a well-studied, kernel-based classification algorithm that searches for a linear decision hyperplane with the largest margin between positive and negative training cases. It is arguably one of the best benchmark classifiers for text categorization. The purpose of our experiments was to assess the effects of semantic-preserving dimension reduction on the classification performance of SVM. We used a publicly available implementation, scikit-learn~\cite{scikit-learn}. We used the linear kernel with $l_2$ regularization and only adjusted the cost-factor parameter, which specifies how the cost of an error on a positive training case outweighs that of a negative case.

\begin{table}
\centering
\caption{Paired Student T-test}
\label{tab:testing}
\begin{tabular}{ll|ll}
\cline{1-4}
\multicolumn{2}{c}{RT}           & \multicolumn{2}{c}{EOD}          \\
\cline{1-4}
Baseline & SVM  & Baseline & SVM \\
0.728           & 0.728          & 0.769           & 0.743          \\
0.728           & 0.728          & 0.769           & 0.769          \\
0.752           & 0.741          & 0.769           & 0.769          \\
0.752           & 0.731          & 0.769           & 0.769          \\
0.752           & 0.752          & 0.757           & 0.757          \\
0.752           & 0.752          & 0.757           & 0.757          \\
0.747           & 0.747          & 0.757           & 0.757          \\
0.747           & 0.747          & 0.757           & 0.757          \\
0.747           & 0.780          & 0.757           & 0.729          \\
0.747           & 0.795          & 0.827           & 0.827          \\
\cline{1-4}
\multicolumn{2}{l}{pvalue=0.471} & \multicolumn{2}{l}{pvalue=0.168}
\end{tabular}
\end{table}

\section{Results and Discussion}
\begin{figure}
\centering
\includegraphics[width = 0.49\textwidth]{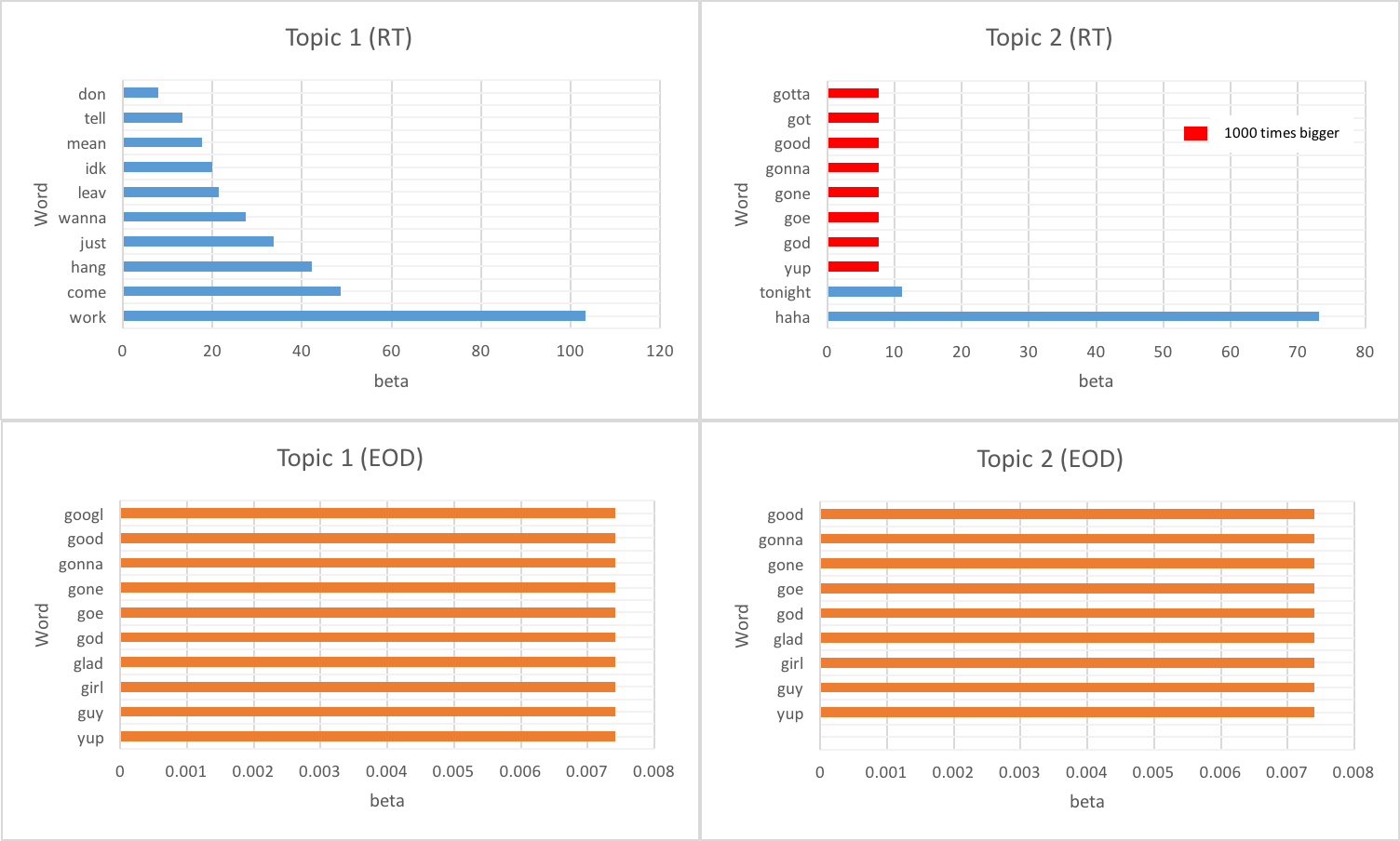}
\caption{Top 2 Topics Learned}
\label{fig:topics_graph}
\end{figure}

Table \ref{tab:results} presents the results from our experiments, averaged over 10-fold cross validation for model selection. We considered several techniques for feature representation namely; the uni-gram, bi-gram and topics (for semantically enriched representation). We used the bi-gram to improve the performance on the uni-gram model, though the gain was marginal$~(\sim +0.02)$ for both RT and EOD. We also explored the techniques of reducing the feature dimension using topic modeling as done in ~\citep{lu2006enhancing}. This approach has the merit of improving model generalization and explanation since meaning can be directly inferred from the most relevant topics , see Figure \ref{fig:topics_graph}. We use the implementation from ~\citep{scikit-learn}, and iteratively select the number of topics over a range of $[0, 200]$ with a step size of 5. We choose the number of topic with the highest F1 score since this metric balances the results from precision and recall, and thus suitable for problem with class imbalance~(e.g., the loneliness rate are 17.52\% and 15.85\%, for RT and EOD respectively). The optimal number of topics are 130 and  135, for RT and EOD respectively. While this approach does not improve the performance over the bag-of-words models, it certainly provide a lot of insights into the predictivness of topics or mixture distribution of words in the model.

\begin{figure}
\centering
\includegraphics[width = 0.49\textwidth]{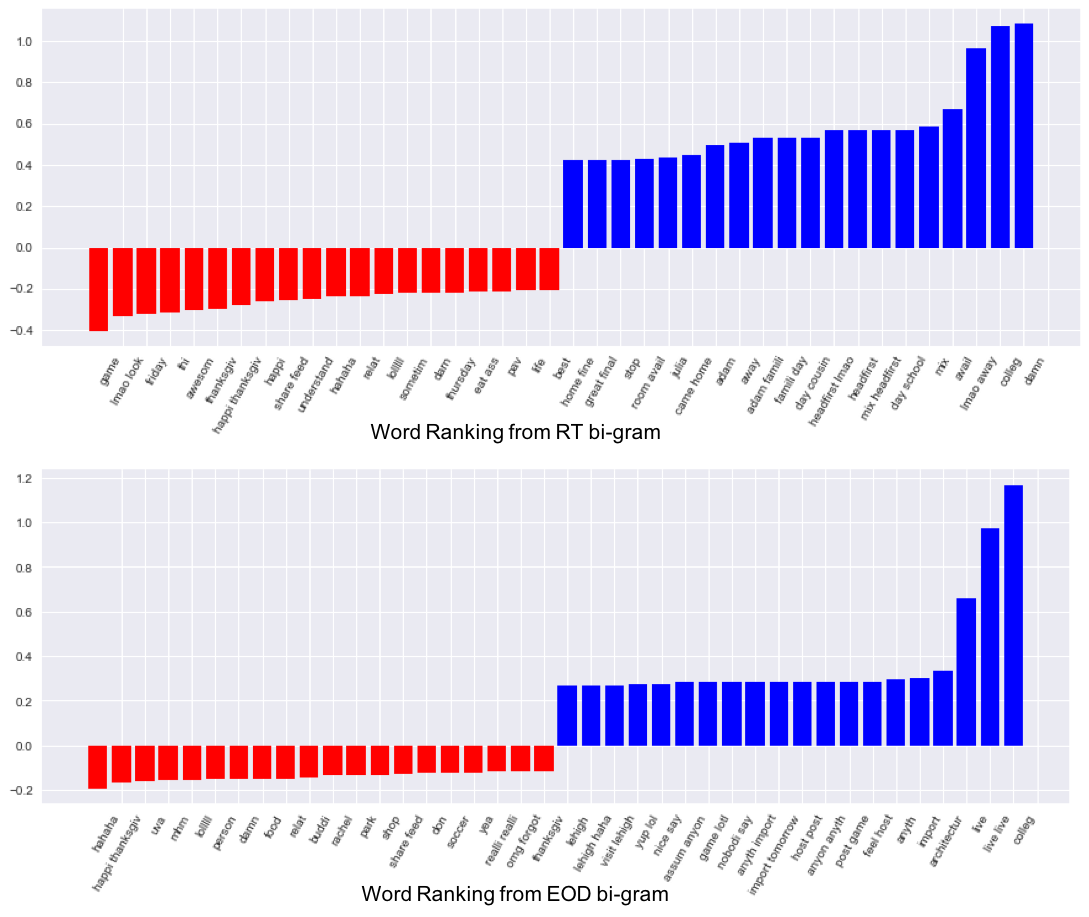}
\caption{Word Ranking from RT/EOD bi-gram}
\label{fig:word_rank}
\end{figure}

In addition to the models explored for the prediction task, we estimated the utility of these models as compared with a baseline rule where an agent predicts the majority class~(not lonely) 100\% of the time. We used the paired t-test over the 10 fold cross validation using F-1 score fro RT and EOD. The results of the test are presented in table~\ref{tab:testing}. With an mean values of 0.745 and 0.769 for baselines of RT and EOD respectively, the t-test  results shows that at $\alpha  = 5\%$ both results are not equal thus the SVM for RT shows significant performance over the baseline while the baseline is better for EOD. This results is further corroborated by the rank of words from the SVM of RT and compared to that of EOD in Figure~\ref{fig:word_rank}. Indeed, while we might expect words like ``damn'' and ``available'' to positively correlate with loneliness, it is unclear how words like ``bye bye'' or ``architecteur'' will positively correlate with loneliness. Also, from the word mixture distributions from the top 2 topics extracted from each model in Figure~\ref{fig:topics_graph}, it is evident that words like ``come'', ``hang'' and even ``work'' might be telling of a person's loneliness in the RT topics, while for the EOD, the $\beta$ parameters from the Dirichlet distribution is equal across all words indication a situation of uncertainty or ``chaos''.

\begin{table*}
 \centering
  \begin{tabular}{cccccccccc}
    \cline{1-10}
    & & \multicolumn{8}{c}{Performance Metrics} \\
    & & \multicolumn{2}{c}{Accuracy} & \multicolumn{2}{c}{Precision} & \multicolumn{2}{c}{Recall} & \multicolumn{2}{c}{F-1} \\
    Data Type & Features & M & CL & M & CL & M & CL & M & CL\\
    \hline
    \multirow{3}{*}{EOD}  & \multicolumn{1}{|l}{Uni-gram} & \multicolumn{1}{|l}{0.81} & 0.04 & \multicolumn{1}{|l}{0.70} & 0.02 & \multicolumn{1}{|l}{0.81} & 0.04 & \multicolumn{1}{|l}{0.75} & 0.02  \\
     & \multicolumn{1}{|l}{Bi-gram} & \multicolumn{1}{|l}{0.83} & 0.02 & \multicolumn{1}{|l}{0.71} & 0.02 & \multicolumn{1}{|l}{0.83} & 0.02 & \multicolumn{1}{|l}{0.76} & 0.02  \\
    & \multicolumn{1}{|l}{Topic-based} & \multicolumn{1}{|l}{0.64} & 0.13 & \multicolumn{1}{|l}{0.76} & 0.08 & \multicolumn{1}{|l}{0.64} & 0.14 & \multicolumn{1}{|l}{0.66} & 0.13  \\
 \multirow{3}{*}{RT} & \multicolumn{1}{|l}{Uni-gram} & \multicolumn{1}{|l}{0.80} & 0.02 & \multicolumn{1}{|l}{0.71} & 0.04 & \multicolumn{1}{|l}{0.80} & 0.02 & \multicolumn{1}{|l}{0.74} & 0.02 \\
     & \multicolumn{1}{|l}{Bi-gram} & \multicolumn{1}{|l}{0.82} & 0.01 & \multicolumn{1}{|l}{0.71} & 0.05 & \multicolumn{1}{|l}{0.82} & 0.01 & \multicolumn{1}{|l}{0.75} & 0.02  \\
     & \multicolumn{1}{|l}{Topic-based} & \multicolumn{1}{|l}{0.62} & 0.09 & \multicolumn{1}{|l}{0.74} & 0.04 & \multicolumn{1}{|l}{0.62} & 0.09 & \multicolumn{1}{|l}{0.66} & 0.08 \\
    \hline
  \end{tabular}
  \caption{Results for classification of Lonely versus not Lonely. M = mean. CL = Confidence Level.}
  \label{tab:results}
\end{table*}

\section{Limitation and Future work}
This findings are limited in a variety of ways; first, the duration of the study~(two weeks) is not enough to sufficiently capture the language pattern as it relates to the person's loneliness level. Methods like data augmentation and semi-supervised learning~\citep{zhu2005semi} in combination with deep learning~\cite{adewole2020deep} can be explored in future work to address this problem. Also, the different subgroups in a cohort might might express varied levels of loneliness, as a result, using clustering approaches~\cite{ameko2018cluster} might reveal these patterns. It is worth noting that the labels though assumed to be ground truth for the purpose of analysis are inherently undermined by recall bias. The effect of this is more pronounced in the EOD model since EOD is meant to assess a person's overall loneliness in the day, much less so for the RT label as they happen more frequently. Finally the ability of the model to generalize beyond the cohort of study is questionable as different age groups communicate in different ways, plus the study cohort is mainly students from Psychology department.

\section{Conclusion}
The study of loneliness is a worthy pursuit because of the health and social implications it has on an individual. Previous studies have been inconclusive on the association of this phenomenon with the use of social media. Specifically the topic of investigating language patterns to predict loneliness has been unexplored in the literature. This work represents a first attempt at this challenging problem from a data mining perspective. Our results, though not outstanding, give a flicker of hope in this direction. More specifically, we have empirically showed that by capturing EMA data at a granular level might improve performance and doing so in a principled manner~(e.g., active learning or optimal experimental design) will more understanding in language patterns as it relates to loneliness.

\bibliography{anthology,emnlp2020}
\bibliographystyle{acl_natbib}

\end{document}